# Pressure Effect on Superconductivity of Iron-based Arsenic-oxide ReFeAsO$_{0.85}$ (Re=Sm and Nd)

Wei Yi, Liling Sun[*], Zhian Ren, Wei Lu, Xiaoli Dong, Hai-jun Zhang，Xi Dai，
Zhong Fang, Zhengcai Li, Guangcan Che, Jie Yang, Xiaoli Shen,
Fang Zhou，Zhongxian Zhao[*]

Institute of Physics and Beijing National Laboratory for Condensed Matter Physics,
Chinese Academy of Sciences, Beijing 100190, P. R. China

Here we report pressure effect on superconducting transition temperature (Tc) of ReFeAsO$_{0.85}$ (Re= Sm and Nd) system without fluorine doping. *In-situ* measurements under high pressure showed that Tc of the two compounds decrease monotonously over the pressure range investigated. The pressure coefficients dTc/dP in SmFeAsO$_{0.85}$ and NdFeAsO$_{0.85}$ were different, revealing the important influence of the deformation in layers on Tc. Theoretical calculations suggested that the electron density of states decreases with increasing pressure, following the same trend of experimental data.

Corresponding author:
llsun@aphy.iphy.ac.cn
zhxzhao@aphy.iphy.ac.cn

The recent discovery of superconductivity with transition temperature of 26 K in iron-based compound La($O_{1-x}F_x$)FeAs [1] has attracted considerable attentions in scientific community. The main reason for such a strong impact is that this high Tc compound contains iron which seems to participate in superconductivity. X-ray diffraction measurements with a synchrotron radiation source show that the compound has the tetragonal crystal structure with space group of P4/nmm at ambient pressure [2]. Theoretical calculations indicate that the electronic structure of La($O_{1-x}F_x$)FeAs is quasi-two dimensional and similar to typical semi-metal [3-9]. By substituting the tri-valence La by rare-earth elements Ce, Pr, Nd, Sm, Gd, the critical transition temperature (Tc) increases significantly and establishes the record of 55 K at ambient pressure up to date [10-15]. Resistance and DC magnetization measurements demonstrate that high pressure in the range of 0-4 GPa displays a positive effect on Tc of La($O_{1-x}F_x$)FeAs [16-18]. The maximum Tc reached 43 K at ~ 4 GPa [17]. In the same pressure range, experimental results for Sm($O_{1-x}F_x$)FeAs showed that Tc strongly relied on the doping level of elemental fluorine [19-20].

Generally, it is thought that fluorine doping plays an important role for the Tc enhancement in iron-based superconductors due to that the substitution of $O^{2-}$ by $F^-$ can induce more electrons into Fe-As layers. Recently, we found that introducing oxygen vacancies to ReFeAsO$_{1-\delta}$ system (Re=La, Ce, Pr, Nd, Sm and Gd ) can also enhance the Tc [12,13]. Moreover, the lattice parameter *a* dependence of Tc of the iron-based arsenic-oxides with doping from La to Gd exhibits dome-like shape, suggesting that the superconductivity transition temperature in ReFeAsO$_{1-\delta}$ system may be related to the compressibility of each compound. The optimal doping level for the highest Tc in the ReFeAsO$_{1-\delta}$ system is very close to $\delta=0.15$ as determined in Ref. [21]. In this study, we selected two compounds of the ReFeAsO$_{1-\delta}$ system (Re= Sm and Nd) and studied the dependence of Tc on pressure via *in-situ* resistance and magnetization measurements under high pressure. We found that Tc decreases monotonically with increasing pressure imposed in the compounds investigated.

The polycrystalline samples used in this study were synthesized under high pressure and high temperature [12] and have been well characterized by x-ray diffractions. In the present study, we carried out resistance measurements of ReFeAsO$_{0.85}$ samples (Re= Sm and Nd) in a diamond anvil cell. Diamonds with 1mm tip were employed for the first run and 300 μm tip with 8° beveled angle for the second run. Four-standard-probe technique was adopted in the resistance measurements, in which four 2 μm-thick platinum (Pt) leads are insulated from rhenium gasket by a thin layer of the mixture of cubic boron nitride and epoxy. In the first run, a bulk sample cut from the synthesized pellet was placed on the top of the diamond and pressed into insulating layer with four Pt leads. In the second run, powder sample taken from the same synthesized pellet was re-pressed into a flake, with ~250 μm in diameter and ~20 μm in thickness, using glassy slides and thereafter the flake was loaded into the diamond anvil cell. Pressure determined by ruby fluorescence method at room temperature [22] was calibrated through the measurements of Tc change of lead as a function of pressure [23]. The superconductivity transition of the samples at each loading point was detected using

CSW-71 refrigerator system.

The magnetization measurements under hydrostatic pressure up to 1 GPa were carried out using Quantum Design Magnetic Property Measurement System (MPMS-XL1). A commercial pressure cell (Mcell 10) was used to create hydrostatic pressure. The pressure was applied at room temperature and determined against Tc change of metal tin at low temperature.

Fig.1(a) shows representative electrical resistance ratio ($R/R_{80K}$) of the SmFeAsO$_{0.85}$ samples as a function of temperature at different pressures. At ambient pressure, the onset Tc of this sample is 55 K. The DC magnetization measured under zero-field cooling (ZFC) and field-cooling (FC) shows sharp diamagnetic transition, demonstrating genuine bulk superconductivity [12]. With increase of pressure, the $R/R_{80K}$-T curves of the bulk sample shifted to the low temperature. As the bulk sample was crashed at 0.65 GPa, $R/R_{80K}$-T curve lost zero resistance upon further increasing pressure, and the superconducting transition temperature shifted towards lower temperature. Examination by a scanning electron microscopy clarified that the nonzero resistance background can be attributed to micro-cracks introduced during non-hydrostatic pressuring. The pressure dependence of superconducting transition temperature for two independent experiments was plotted in Fig.1(b). The onset superconducting transition temperature (Tco) in this study was determined by the dR/dT. It is obvious that the Tco of the SmFeAsO$_{0.85}$ sample displayed a monotonous decrease with imposed pressure. The fitting to the data yields dTco/dP=2.0 K/GPa. *In-situ* magnetization measurements showed the same trend that both Tco and Tpeak (temperature at the peak of dM/dT, where the slope of superconducting transition reaches its maximum) of the sample went down with pressure increase, as shown in Fig.2. After decompression from high pressure the Tco values are slightly hysteretic than that detected from uploading, but basically follow a linear behavior, indicating that the pressure effect on Tco in this compound is reversible. The fittings to the dTco/dP as well as dTpeak/dP data obtained from magnetization measurements under hydrostatic pressure up to 1 GPa in $SmFeAsO_{0.85}$ result in pressure coefficients of dTco/dP = -1.3 K/GPa and dTpeak/dP = -2.0 K/GPa respectively, which are consistent with the tendency of dTco/dP value (-2.0 K/GPa) achieved by resistance measurements under nonhydrostatic pressure.

The superconducting transition temperature of $NdFeAsO_{0.85}$ as a function of pressure was also studied. At ambient pressure, the superconducting $NdFeAsO_{0.85}$ sample held an onset Tc at 51.9 K which is a few Kelvin lower than the Tc value (55 K) of $SmFeAsO_{0.85}$ [12]. According to the report of Ref. [12], the lattice parameter *a* of $NdFeAsO_{0.85}$ (0.395 nm) is larger than that of $SmFeAsO_{0.85}$ (0.390 nm), so that Tc of $NdFeAsO_{0.85}$ was expected to be increased under high pressure. But our results show that the Tc exhibits a monotonous decrease as pressure increases. Fig.3 shows the sample resistance as a function of temperature and the pressure dependence of Tc for the $NdFeAsO_{0.85}$ sample. During the first round of review process of this paper, we were aware of the pressure study report for $NdFeAsO_{0.6}$ [24] and realized that the tendency of pressure effect on Nd-based sample without fluorine doping is almost the same as what we observed. It therefore suggests that the Tc of the compounds with optimal doping should be decreased under pressure, despite of the reduction of lattice parameter $a$. Comparing the difference of pressure effects on SmFeAsO$_{0.85}$ and the NdFeAsO$_{0.85}$, we found that Tco of NdFeAsO$_{0.85}$ is more sensitive to the pressure than that of SmFeAsO$_{0.85}$. The dTco/dP is nearly -2.6 K/GPa for NdFeAsO$_{0.85}$ and -2.0 K/GPa for SmFeAsO$_{0.85}$, as shown in Fig.4 (a).

By first-principle calculations based on density functional theory, we have calculated the electron density of states (DOS) of the system as a function of pressure. The crystal structures are fully optimized and the calculated results are shown in Fig.4(b). It is clear that the electron density of states decreases with the increasing pressure, following the same trend of Tc as function of pressure obtained from the current high pressure experiments. In generalized BCS theory, the critical temperature is determined by both the density of states at the Fermi level and the strength of the pairing interaction. The present results may suggest that the latter depends on the pressure very weakly for the systems we studied here.

In summary, a universal Tc decrease was observed in optimal oxygen doped SmFeAsO$_{0.85}$ and NdFeAsO$_{0.85}$ samples under high pressure. The pressure coefficients dTc/dP in SmFeAsO$_{0.85}$ and Nd FeAsO$_{0.85}$ were different. The Tco of NdFeAsO$_{0.85}$ is more sensitive to the pressure than that of SmFeAsO$_{0.85}$. The common origin for negative pressure effect on Tc is that pressure drives a decrease of the electron density of states at Fermi level. A complete understanding of these observations requires further investigation

Acknowledgments

We sincerely thank Prof. L. Yu for valuable discussions. We wish to thank the National Science Foundation of China for its support of this research through Grant No. 50571111 and 10734120. This work was also supported by the Ministry of Science and Technology of China (2005CB724400, 2006CB601001 and 2007CB925002). We also acknowledge the support from EU under the project CoMePhS.

Figure and captions:

Fig.1 (a) Representative electrical resistance of the $SmFeAsO_{0.85}$ samples as a function of temperature at different pressures. (b) Pressure dependence of Tc of the $SmFeAsO_{0.85}$ sample for different runs. The onset Tc (Tco) was defined by dR/dT.

Fig.2 Left panel presents the differential curves of diamagnetic magnetization (dM/dT) measured under 10 Oe after FC (field cooling) under different hydrostatic pressures. The pressure is labeled from the top down in the order of experimental runs. Right panel shows the pressure dependence of onset Tc as well as the peak of dM/dT derived from left panel, solid lines are guides for eye.

Fig.3 (a) Resistance-temperature curves of the $NdFeAsO_{0.85}$ samples at different pressures and (b) onset superconducting transition temperature Tco as a function of pressure.

Fig.4 (a) Comparison of the pressure shift of Tco in the SmFeAsO$_{0.85}$ and the NdFeAsO$_{0.85}$ sample, showing a larger negative pressure effect on the Tco in NdFeAsO$_{0.85}$ sample. (b) Calculated pressure dependence of electron density of state from first-principle calculations.

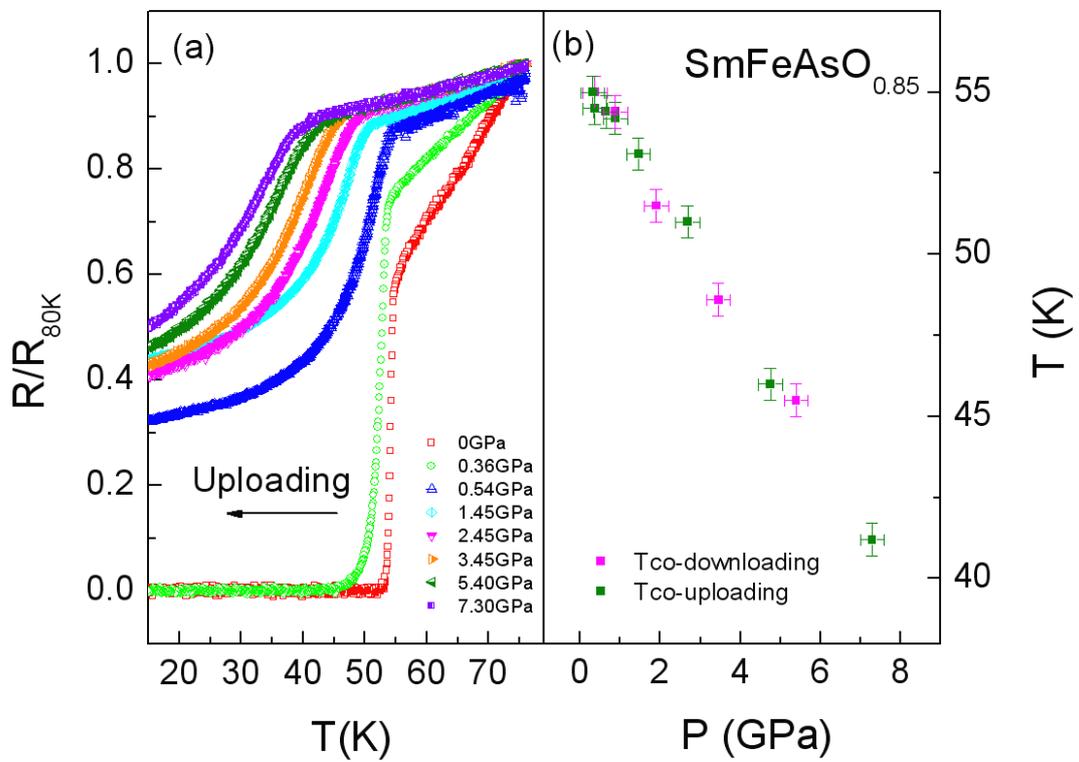

Fig.1

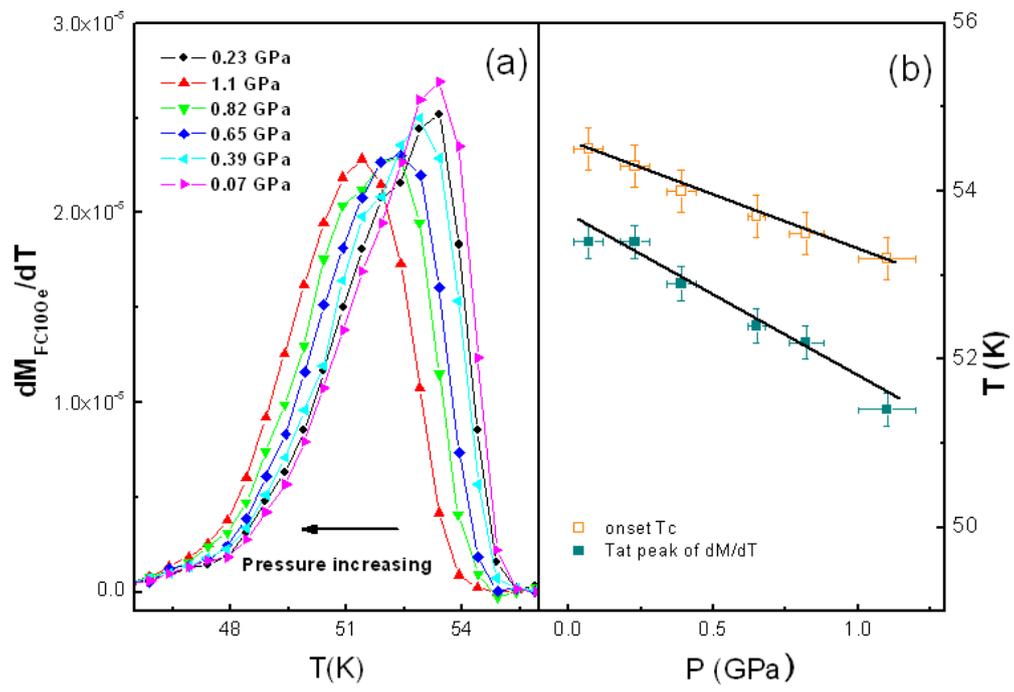

Fig.2

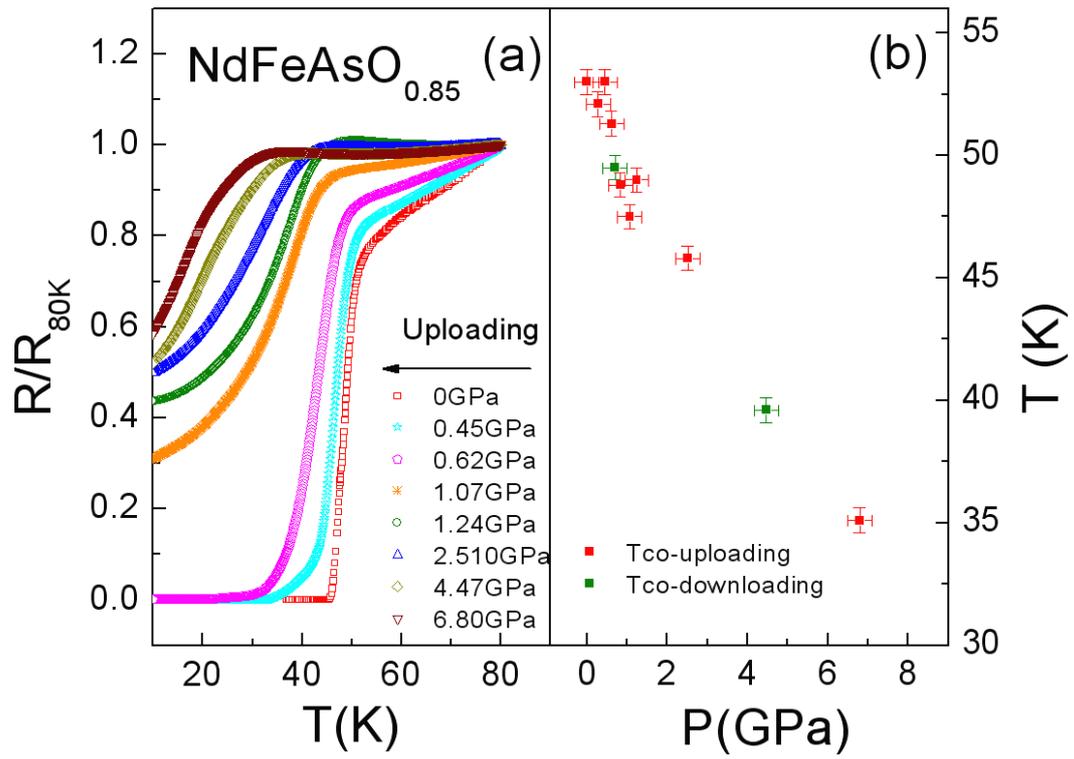

Fig. 3

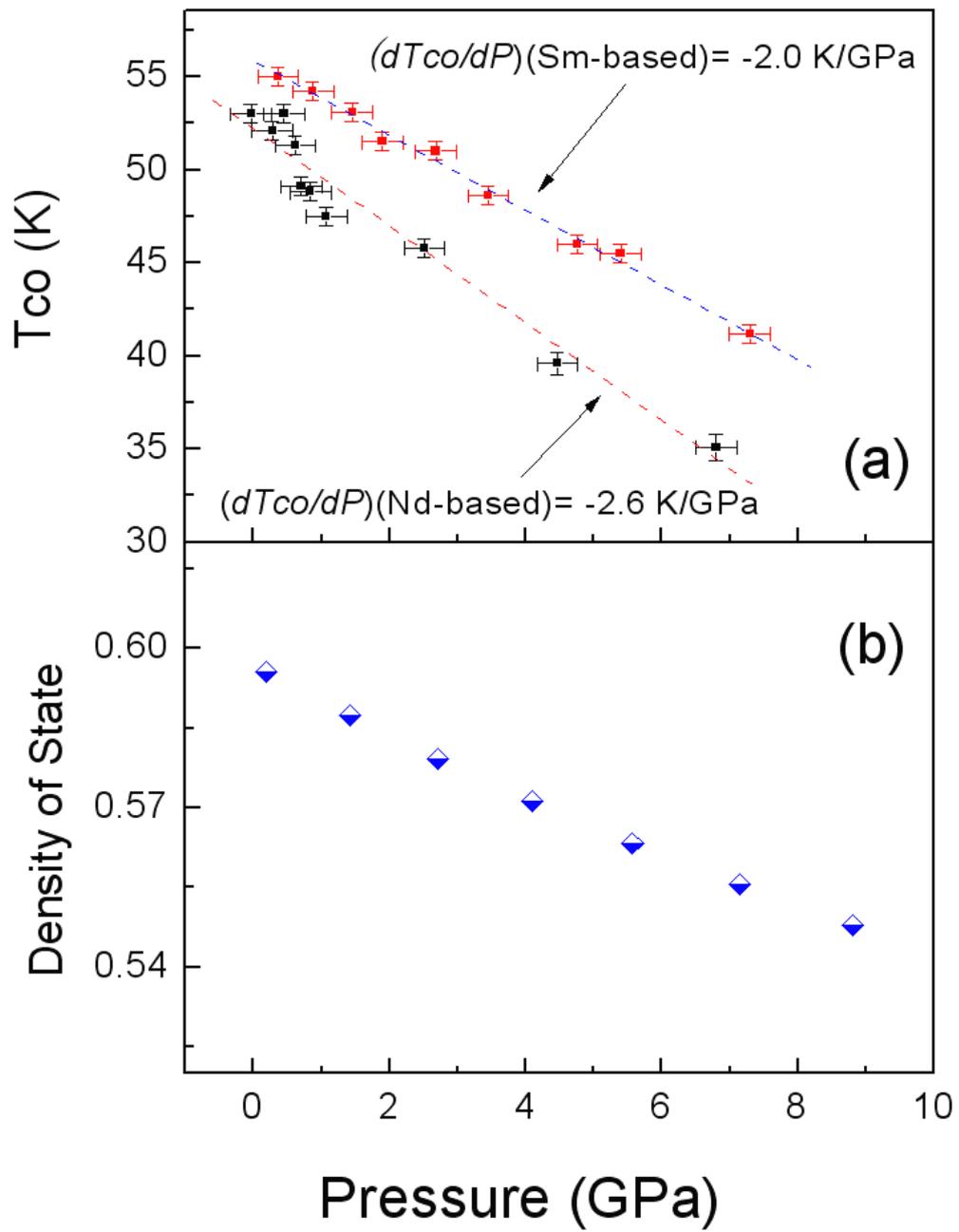

Fig.4